\documentclass[pra,superscriptaddress,twocolumn,showpacs]{revtex4}
\usepackage{amssymb}
\usepackage{amsmath}
\usepackage{color}
\usepackage{dsfont}
\usepackage{hyperref}
\usepackage{graphicx}

\definecolor{darkblue}{rgb}{0,0,0.5}
\definecolor{darkred}{rgb}{0.5,0,0}

\hypersetup{
  pdfborder={0,0,0},
  colorlinks=true,
  linkcolor=darkblue,
  citecolor=blue
}

\begin{document}

\title{Interferometry with relativistic electrons}

\author{K.-P. Marzlin}
\affiliation{Department of Physics, St. Francis Xavier University,
  Antigonish, Nova Scotia, B2G 2W5, Canada}
\affiliation{Institute for Quantum Information Science,
        University of Calgary, Calgary, Alberta T2N 1N4, Canada}

\author{T. Lee}
\affiliation{Department of Physics, St. Francis Xavier University,
  Antigonish, Nova Scotia, B2G 2W5, Canada}

\bigskip

\begin{abstract}
We propose an experiment to test the influence of Lorentz contraction
on the interference pattern of a beam of electrons. 
The electron beam is split and recombined by two pairs of bi-chromatic laser
pulses, using a variation of the Kapitza-Dirac effect. Between the
pairs the electrons are accelerated to relativistic speed.
We show that Lorentz contraction of the distance between two
partial beams will then lead to a reduction of fringe visibility.
The connection of the proposal to Bell's spaceship paradox is discussed.
\end{abstract}



\maketitle

\section{Introduction}
The influence of Special Relativity on matter-wave interference 
is a long-established fact. For instance, the spatial interference
pattern of atom interferometers that are based on 
optical Ramsey fringes \cite{Borde198910} 
can only be fully explained if the relativistic Doppler effect is 
taken into account \cite{PhysRevA.30.1836}. 
This effect is relevant even for non-relativistic velocities because 
the fringe pattern is very sensitive to differences between the dynamical phase
factors $\exp (-iE t/\hbar)$ of different partial beams in the interferometer.
The relativistic Doppler effect essentially takes into account 
time dilation in the dynamical phase.

On the other hand, the consequences of Lorentz contraction in
matter-wave interference are much more difficult to detect.
Lorentz contraction generally has only been confirmed indirectly, for instance
through the compressed charge distribution in high-energy 
ion collisions \cite{PhysRevD.27.140}
and the wavelength of free-electron lasers \cite{Margaritondo:kv5089}.
The Michelson-Morley experiment, which was the reason for
the introduction of Lorentz contraction, may also be considered as an indirect
confirmation. An overview about experimental tests of Lorentz contraction
can be found in Ref.~\cite{physics0603267}.

In this paper we suggest an experiment to observe the impact of  
Lorentz contraction on the spatial interference pattern of an
electron interferometer. The principle idea of the proposal is that an
electron beam is split, accelerated to relativistic speed and then
recombined. Lorentz contraction of the distance between two partial
beams will then lead to a reduction of fringe visibility. In
Sec.~\ref{sec:expSketch} we will outline how to realize this scheme
using a modification of the Kapitza-Dirac effect. A theoretical
analysis of the interference pattern based on the Dirac equation
in Sec.~\ref{sec:theory} is followed by
an analysis of the beam splitting process in Sec.~\ref{Sec:BS}
Numerical results for the interference
pattern are presented in In Sec.~\ref{sec:Numeric}, and in
Sec.~\ref{sec:Bell} the connection of the proposal to Bell's
spaceship paradox is discussed.
\begin{figure}
\begin{center}
a) \includegraphics[height=4.5cm]{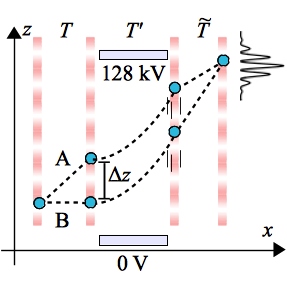}\hspace{5mm}
b) \includegraphics[height=4.5cm]{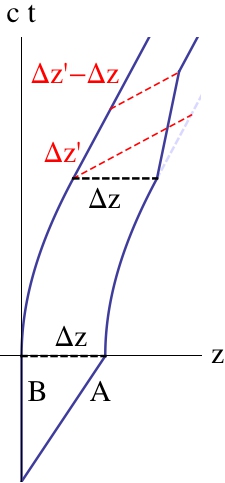}
\caption{\label{fig:expSketch}
a) Sketch of the proposed experiment. An electron beam is
split and recombined using four bi-chromatic laser pulses
and accelerated to relativistic speed. Lorentz contraction of the
beam separation reduces fringe visibility.
b) Space-time diagram of the split electron beam.
}
\end{center}
\end{figure}

\section{Sketch of proposed experiment}\label{sec:expSketch}
Our proposal employs the general principles of 
Ramsey-Bord\'e interferometers
\cite{Borde198910,PhysRevLett.67.177,PRA47:4441,PRA50:2080}, in which
an atomic beam is split and recombined by a sequence of laser
pulses. When passing through a laser pulse the atoms absorb photons,
so that momentum is transferred from the pulse to the atoms. This
change in the atomic center-of-mass dynamics can be used to construct
atom beam splitters. In a similar way, the Kapitza-Dirac 
effect
\cite{PSP:1736920,RevModPhys.79.929,PhysRevLett.61.1182,Nature413-142}
can be used to transfer momentum from a standing light wave to electrons.
A Ramsey-Bord\'e interferometer for (non-relativistic) electrons that employs
bi-chromatic laser pulses as beam splitters has 
been described in Ref.~\cite{KPM-eInterferometer2013}. Our proposal
builds on this work.

A sketch of the suggested experiment
is shown in Fig.~\ref{fig:expSketch} a).
An electron beam initially
moves in the $x$-direction and is then coherently split
into two beams A, B that are spatially separated by a distance $\Delta z$. 
The splitting is accomplished by two bi-chromatic laser pulses, which
are represented as dashed vertical red lines in  Fig.~\ref{fig:expSketch} a).
The first pulse (left-most dashed line) splits
the electron beam into two partial beams and transfers
momentum to one of the beams, thus changing their relative velocity
$\Delta v$. Between the first two pulses electrons travel 
freely for a time $T$ so that they acquire a distance $\Delta z = T
\Delta v$. The second laser pulse reverses the momentum transfer so
that beam A and B have the same momentum.

The split electron beam then enters a region with a strong electric field that accelerates
it in the $z$-direction for a time $T'$.
After the electrons have passed through the electric field and
obtained a relativistic speed $v=\beta c$, the electrons are recombined
in such a way that the spatial distance between the two beams 
is reduced by an amount $\Delta z$
in their rest frame. For non-relativistic electrons this would 
lead to a perfect
overlap, resulting in an interference pattern with high fringe
visibility, but at relativistic speed Lorentz contraction changes this
conclusion.

To understand this, consider the space-time diagram of the
interferometer in Fig.~\ref{fig:expSketch}b), which shows 
that, in the laboratory frame, the distance between beam A and B
remains unchanged during the acceleration. However, once the
electrons are moving at speed $v$, the proper distance $\Delta z'$ in their
rest frame has to be measured simultaneously in that frame, 
i.e., along the lower red dashed line in Fig.~\ref{fig:expSketch}b)
(see also Sec.~\ref{sec:Bell}).
We then have $\Delta z' = \gamma \Delta z > \Delta z$, where
$\gamma = 1/\sqrt{1-\beta^2}$, because $\Delta z$ is the Lorentz
contraction of $\Delta z'$. Consequently, after the recombination
beam A and B  would miss each other
by a distance $\Delta z' -\Delta z= (\gamma-1)\Delta z$, which would
lead to reduced fringe visibility.

\section{Theoretical analysis of the interferometer}\label{sec:theory}
To estimate the achievable magnitude of the beam separation $\Delta z' -\Delta z$
we extend the description of a Ramsey-Bord\'e interferometer for electrons
\cite{KPM-eInterferometer2013} to the relativistic regime.
It is assumed that the beam intensity is low enough so that
space-charge effects can be neglected \cite{Freimund:Thesis2003}.
We first summarize the key points of 
the non-relativistic description.
\\
(i) The initial electron state is expanded in terms
of momentum eigenstates $\phi_{p}(z) $ as
\begin{align} 
   \psi_{\text{init}}(z) &=\int dp \,
   \tilde{\psi}(p) 
   \phi_{p}(z) .
\label{eq:initState}\end{align} 
It is sufficient to describe the motion along the $z$-axis
because all forces point along this direction.
We consider states for which $\tilde{\psi}(p) $ is negligible
if not $|p|\ll \hbar k_L$, with $k_L$ the wavenumber 
of the laser pulses.
\\
(ii) The free evolution of the electrons for a time $T$ 
between two laser pulses amounts to the
replacement of $\tilde{\psi}(p) $ by
$\exp [-iTE(p)/\hbar]\, \tilde{\psi}(p) $, where $E(p)$ is the
energy of an electron with momentum $p$.
\\
(iii) Acceleration for a time $T'$ amounts to a
unitary transformation $\hat{U}_a(T') $ of the form
\begin{align} 
  \hat{U}_a(T') \phi_{p}(z) &= e^{-i \tau(p)} \;
  \phi_{p+m a T'}(z),
\label{eq:schroedSolAccel}
\\
   \tau(p) &= \frac{ 1}{\hbar} \int_0^{T'} dt'\, E(p+m a t').
\label{eq:tauDef}\end{align} 
(iv) The first and second laser
pulse induce a unitary transformation
\begin{align} 
    \hat{U}_{-} \phi_{p}(z)&=
   \frac{ 1}{\sqrt{2}} \left (
  \phi_{p}(z) + \phi_{p+2\hbar k_L}(z)
  \right )
\\ 
     \hat{U}_{-} \phi_{p+2\hbar k_L}(z)&=
   \frac{ 1}{\sqrt{2}} \left (
  -\phi_{p}(z) + \phi_{p-2\hbar k_L}(z)
  \right ).
\label{eq:UtrafoSimpl}\end{align} 
The last two laser pulses produce a similar unitary transformation
$\hat{U}_{+}  $, which is equal to $\hat{U}_-$ with $k_L$ replaced by $-\tilde{k}_L$. 
These transformations are not accurate but provide a reasonable
approximation for non-relativistic electrons. We will 
give a detailed description of the splitting process below.
\\
(v) Using point (i)-(iv), the electron state after passing through
the interferometer evaluates to \cite{KPM-eInterferometer2013}
\begin{align} 
  \psi&_{\text{final}}(z) = 
  \frac{ 1}{4} \int dp \,  \tilde{\psi}(p)     \phi_{p+maT'}(z) 
\nonumber \\ & \hspace{0cm}\times 
   e^{-i \tau(p,T')}   \Big (
  e^{-\frac{ i T}{\hbar} E (p+2\hbar k_L) }  
  e^{-\frac{ i \tilde{T}}{\hbar} E(p-2\hbar \gamma \tilde{k}_L+m a T') }
\nonumber \\ & 
   + e^{-\frac{ i T}{\hbar} E(p) }  
  e^{-\frac{ i \tilde{T}}{\hbar} E(p+m a T') }
\Big ) + {\mbox{rest}}.
\label{eq:finalStateNonrel}\end{align} 
The first term in parentheses corresponds to beam A
in Fig.~\ref{fig:expSketch} a), i.e., to electrons which receive
momentum transfers $2\hbar k_L, -2\hbar k_L, -2\hbar \tilde{k}_L, 2\hbar \tilde{k}_L$
at the four laser pulses. The second term in parentheses corresponds
beam B, with electrons that travel
through the laser pulses without changing their momentum.
``Rest'' refers to (seven) other partial beams that are produced in
addition to beam A and B. For brevity these beams will only be
included in the numerical analysis given below. 

We now adapt this derivation to relativistic electrons
described by the Dirac
equation $i\hbar \partial_t \psi = \hat{H}\psi$, with
\begin{align} 
  \hat{H} &= mc^2 \underline{\beta} +\hat{H}_0 + q V
  -c \sum_{i=1}^3 (\hat{p}_i -q A_i) \underline{\alpha}_i .
\label{eq:DiracHam}\end{align} 
The $4\times 4$ matrices $\underline{\beta}$ and $\underline{\alpha}_i$ take
their standard form \cite{BjorkenDrell}.
The initial state can still be expanded as in Eq.~(\ref{eq:initState})
with $\phi_p(z)$ replaced by spinor momentum eigenstates
$\phi_{p}^{(r)} (z) = \exp(i p z/\hbar)\, \theta ^{(r)} (p)$, 
where 
\begin{align} 
   \theta^{(1)}(p) &=
  \frac{ 1}{\sqrt{2}} 
  \left (   \sqrt{1+\frac{m}{E}} , 0 , \frac{p}{\sqrt{E
   \left(m+E\right)}}  , 0\right ) 
\\
   \theta^{(2)}(p) &=
  \frac{ 1}{\sqrt{2}} 
  \left (    0, \sqrt{1+\frac{m}{E}},
  0, \frac{-p}{\sqrt{E
   \left(m+E\right)}}
  \right ) 
\\
  \theta^{(3)}(p)&=
  \frac{ 1}{\sqrt{2}}
  \left (
   \sqrt{1-\frac{m}{E}} , 0 ,
   -\frac{p}{\sqrt{E \left(E-m\right)}} , 0
 \right ) 
\\
   \theta^{(4)}(p)&=
  \frac{ 1}{\sqrt{2}} 
  \left (  0 , \sqrt{1-\frac{m}{E}}
  , 0 ,  \frac{p}{\sqrt{E  \left(E-m\right)}}
  \right ) ,
\label{eq:planeWaves2}\end{align} 
with the relativistic energy $E(p)=\sqrt{p^2c^2+m^2c^4}$.
Using this expression for $E(p)$, the free evolution rule (ii) can
also be used for relativistic electrons.

To see that the acceleration rule (iii) can still be applied we have to
solve the Dirac equation with constant acceleration $a$
\cite{PhysRevD.22.1935,PRD42:2045,SolvableDirac-EPL1990,PhysRevD.43.3979,PhysRevA.74.032326,1751-8121-40-30-024},
corresponding to $A_i=0$ and $qV=-m a z$ in Eq.~(\ref{eq:DiracHam}). 
To keep a close analogy to the non-relativistic treatment of 
Ref.~\cite{KPM-eInterferometer2013} we expand the wavefunction as
\begin{align} 
 \psi(z,t) &= \sum_{r=1}^4 \tilde{\psi}_r(t) \phi_{p(t)}^{(r)}(z),
\end{align} 
with $p(t)=p+m a t$. Inserting this into the Dirac equation and
exploiting the orthonormality of the spinors $\theta^{(r)}(p)$ we obtain
\begin{align} 
  i\partial_t \tilde{\psi}_r&= \epsilon_r \frac{ E(p(t))}{\hbar} \tilde{\psi}_r
  -i \epsilon_r  (-1)^r \eta(t)\tilde{\psi}_{r+2 \epsilon_r },
\end{align} 
with 
\begin{equation} 
   \eta(t) \equiv \frac{  a}{2c} \left (\frac{ m c^2}{E(p(t))}\right )^2
\end{equation} 
 and $\epsilon_r =1 (-1)$ for $r=1,2$ ($3,4$), respectively.
$\eta(t)$ describes a
coupling between positive- and negative-energy solutions that
is maximal for $E(p(t))\approx m c^2$.
For constant values of $\eta$ and $E(p)$ we find that the
maximal transition probability is given by
$\hbar^2\eta^2/(E(p)^2+\hbar^2\eta^2)$, which is only
significant for extreme accelerations of
$a\approx 10^{30}$ m/s$^2$ or larger. For realistic accelerations
$\eta(t)$ is negligible, so that the solution to the accelerated Dirac equation
can be approximated by 
\begin{equation} 
  \tilde{\psi}_r(t) \approx
  e^{-\frac{ i}{\hbar} \epsilon_r \int_0^t dt' E(p(t')) } \tilde{\psi}_r(0).
\end{equation} 
Consequently, Eq.~(\ref{eq:schroedSolAccel}) 
can also be used to describe the accelerated evolution of
relativistic electron wavepackets.

In Sec.~\ref{Sec:BS} it will be shown that beam splitting rule (iv) 
also applies to relativistic electrons if specific
conditions are met. Hence, with the appropriate replacements,
rules (i)-(iv) and hence
final state (\ref{eq:finalStateNonrel}) are still valid for
relativistic electrons.
Compared to Ref.~\cite{KPM-eInterferometer2013} we have admitted
that the time $\tilde{T}$ between the last two pulses,
and the momentum transfer $2\hbar \tilde{k}_L$ 
in the electron rest frame,
differ from the respective values for the first
two pulses. 
A factor of $\gamma$ appears in Eq.~(\ref{eq:finalStateNonrel}),
which is formulated in the lab frame,
because we have to
perform a Lorentz transformation of the momentum transfer
to obtain the momentum transfer in the lab frame.

In the non-relativistic case it is possible to evaluate
Eq.~(\ref{eq:finalStateNonrel}) analytically for a
Gaussian initial wavepacket of spatial width $w$, for which
\begin{equation} 
  \tilde{\psi}(p) =
  \sqrt{\sqrt{\frac{ 2}{\pi}} \frac{ w}{\hbar}}
  e^{-p^2 w^2/\hbar^2} .
\label{eq:momentumComponents} \end{equation} 
In the relativistic case we can obtain an approximate
solution by exploiting that the momentum width of the wavepackets
is still small after the acceleration. We can therefore 
expand  all exponentials to second order in $p$ and
replace the spinor $\theta^{(r)}(p+m a T')$ by $\theta^{(r)}(m a T')$.
Then all integrals in Eq.~(\ref{eq:finalStateNonrel}) are of Gaussian
form and can be solved analytically, leading to partial beams with
a Gaussian spatial structure.
The final mean positions of the
two partial beams in the lab frame evaluate to
\begin{align} 
  z_{\text{A}} &= \Delta z + \frac{ c^2}{a} (\gamma-1)  
     + \tilde{T}\left ( \beta c -
   \frac{ 1}{\gamma^2} 
    \frac{ 2 \hbar \tilde{k}_L }{m} 
     \right )
\label{eq:z1}\\ 
  z_{\text{B}} &= \frac{ c^2}{a} (\gamma-1)  
     + \tilde{T}\beta c ,
\label{eq:z2}\end{align} 
where we have used
$\Delta z = T\Delta v= 2T\hbar k_L/m$ for the 
separation induced by the first pair of laser pulses.
These mean positions
agree with relativistic trajectories of classical point particles
in the lab frame.
$c^2(\gamma-1)/a$ corresponds to the
distance travelled by the electrons during the acceleration
phase. For $\gamma=5/4$ and an acceleration of $1.8\times 10^{16}$~m/s$^2$,
which corresponds to an electric field strength of $10^5$~V/m,
this distance is about 1.3~m.

Terms proportional to $\tilde{T}$ correspond to the distance
travelled by the partial beams between the last two laser pulses.
$\beta c$ is the velocity of electrons with momentum $p=m \gamma \beta c$,
which have not received a
momentum kick. Electrons that have received a momentum kick 
$\Delta \tilde{p}=-2\hbar \tilde{k}_L$ in their rest frame 
possess a momentum $p+\gamma \Delta \tilde{p}$ in the lab frame.
The relativistic relation between velocity and momentum is given by
$\beta(p) = p/\sqrt{p^2+m^2c^2}$. Expanding
$\beta(p+\gamma \Delta \tilde{p})$ to first order in
$\Delta \tilde{p}$ produces the terms proportional to
$\tilde{T}$ in Eq.~(\ref{eq:z1}).

To realize the proposal presented in Sec.~\ref{sec:expSketch}, 
$\tilde{T}$ has to be chosen in such a way that the distance
between beam A and B is reduced by an amount $\Delta z$ between the last two laser pulses. 
If $\Delta \tilde{v} = 2\hbar \tilde{k}_L/m$ denotes
the relative velocity of the two partial beams in their rest frame, the proper time
needed to cover this distance is given by $\tau =\Delta z/\Delta \tilde{v}$.
In the lab frame, the time between the two pulses must therefore
be chosen as 
$  \tilde{T}=\gamma \tau \; = 
   \gamma \Delta z\, m/(2\hbar \tilde{k}_L)$.
The final distance between the two beams in the lab frame
is then given by
$z_{\text{A}}-z_{\text{B}} = \Delta z(1-\gamma^{-1})$. Lorentz contraction
implies that in the rest frame of the electrons the distance
is then given by $\gamma(z_{\text{A}}-z_{\text{B}})=(\gamma-1)\Delta z$,
which is corresponds to the mismatch discussed in Sec.~\ref{sec:expSketch}.

\section{Kapitza-Dirac beam splitter for relativistic electrons}\label{Sec:BS}
The analysis given in Sec.~\ref{sec:theory} employs rule (iv),
which has been derived for non-relativistic electrons
\cite{KPM-eInterferometer2013} using a modified Kapitza-Dirac effect and is only correct
in the limit of very short laser pulses.
Rule (iv) is sufficient to give a rough description of the interaction
of electrons with the first two laser pulses, but it
needs to be reconsidered for relativistic
electrons interacting with the two laser pulses 
after the acceleration.
The relativistic Kapitza-Dirac effect has been studied in
Ref.~\cite{PhysRevLett.109.043601}. 
To describe the modified relativistic Kapitza-Dirac effect
for each of the four pulses
we need to solve the Dirac equation (\ref{eq:DiracHam})  in the presence of
two counter-propagating laser fields with different frequencies.
The corresponding electromagnetic potentials are given by 
$V=0$ and 
$\vec{ A}=\vec{ \epsilon}( A^{(+)}+A^{(-)})$, where
\begin{align} 
  A^{(+)}&= -\frac{i E_1 e^{i k_1 z-i t \omega _1+i\theta_1}}{4 \omega _1}-\frac{i E_2 e^{-i
   k_2 z-i t \omega _2+i\theta_2}}{4 \omega _2}
\label{eq:KDvectorpot}\end{align} 
is the positive-frequency part of the vector potential 
in the lab frame
and $ A^{(-)}= ( A^{(+)})^*$.
$E_i$, and $\omega_i$ ($i=1,2$) are electric field amplitude and frequency of the 
two counter-propagating fields and $k_i=\omega_i/c$ their wavenumber.
The unit vector $\vec{ \epsilon}$ describes the polarization
direction in the $x$-$y$-plane and $\theta_i$ are phase factors.

For non-relativistic velocities
the Dirac equation can be solved using a Foldy-Wouthuysen 
transformation \cite{BjorkenDrell}. The 
large components of the Dirac spinor are then of the form
$\psi=\exp(-i m c^2 t/\hbar) \tilde{\psi}$, where $\tilde{\psi}$
is a solution to the Schr\"odinger equation with
the same vector potential.
The analysis of Ref.~\cite{KPM-eInterferometer2013} can therefore
be applied to the first two laser pulses.

To describe the last two laser pulses we can exploit that, despite
the relativistic
mean electron velocity,
the velocity spread is still non-relativistic. We therefore can
perform a Lorentz boost of the Dirac equation along the $z$-axis into 
the (mean) electron rest frame. Because of the
covariance of the Dirac equation, the result will still be of the form
(\ref{eq:DiracHam}) but with a transformed vector potential.

The Lorentz transformation of the four-vector potential $(0, \vec{
  A})$ can be easily accomplished by noting that its polarization
is perpendicular to the direction of the boost. Consequently,
the vector potential still has the form (\ref{eq:KDvectorpot})
except that in the exponentials we have to make the
replacements $\omega_1 \rightarrow \tilde{\omega}_1\equiv(1-\beta)\gamma \omega_1$ and
$\omega_2 \rightarrow \tilde{\omega}_2\equiv(1+\beta)\gamma \omega_2$. 
Therefore, detuning $\Delta\omega \equiv \omega_2-\omega_1$ and 
average wavenumber $k_L\equiv (k_1+k_2)/2$ in the lab frame
need to be replaced by the respective values in the electron rest frame, 
\begin{align} 
  \Delta\tilde{\omega} &= 
  \gamma\left ( \Delta\omega + \beta ( \omega_2+
  \omega_1)\right )
\label{eq:omgaTilde} \\
  \tilde{k}_L &= \frac{ \gamma}{2c}\left (
    (1+\beta) \omega_2 +
    (1-\beta) \omega_1
  \right ).
\label{eq:kTilde}\end{align} 

It was shown in Ref.~\cite{KPM-eInterferometer2013} 
that rule (iv) provides a reasonable approximation for the evolution
of the electron state inside a bi-chromatic laser pulse if the resonance
condition 
\begin{equation} 
  \Delta\tilde{\omega} = \mp 2\frac{\hbar}{m} \tilde{k}_L^2
\label{eq:resCond}\end{equation} 
is fulfilled. For relativistic electrons
this poses a practical limitation: because of the Doppler effect 
(terms proportional to $\beta$ in Eq.~(\ref{eq:omgaTilde})), the 
detuning $ \Delta\tilde{\omega}$ in the electron rest frame may be
very large. However, the Kapitza-Dirac effect
requires phase locking, i.e., there must be a stable relation between the phases of the two
counter-propagating laser fields \cite{Nature413-142}.
In current experiments such a relation can only be established for small detunings
\cite{0957-0233-20-5-055302} 
or in harmonic generation \cite{PhysRevLett.79.1006}.
We therefore propose the following setup:
laser field 2 with frequency $\omega_2$ is detuned by a small amount $\delta \omega$ from,
and phase-locked to, a pump laser of frequency
$\omega_2-\delta\omega$. Laser field 1 corresponds
to the $n$th harmonic frequency of the pump laser, so that
$\omega_1 = n(\omega_2 -\delta\omega)$.
We then obtain
\begin{align} 
  \Delta\tilde{\omega} &= 
  n (1-\beta) \gamma \delta\omega + 
   (1+\beta -n(1-\beta))
   \gamma \omega_2
\label{eq:omgaTilde2} \\
  \tilde{k}_L &= \frac{ \gamma}{2c}\left [
    (1+\beta + n (1-\beta)) \omega_2 -
    (1-\beta) \delta\omega
  \right ].
\label{eq:kTilde2}\end{align} 
If the final velocity of the electrons takes the value
$\beta=(n-1)/(n+1)$
then these relations simplify to
$ \Delta\tilde{\omega}=\sqrt{n} \delta\omega$ and
$\tilde{k}_L \approx \sqrt{n} \omega_2/c=\sqrt{n}k_2$.
The resonance condition can then easily be fulfilled
by choosing $\delta\omega = \mp 2\hbar \sqrt{n} k_2^2/m$,
which apart from a factor of $\sqrt{n}$ is the same condition
as in the non-relativistic case. 
Hence, if the electrons are accelerated to a specific velocity,
rule (iv) can still be used to describe
the beam splitters. 

\section{Numerical results}\label{sec:Numeric}
\begin{figure}
\begin{center}
\includegraphics[width=7cm]{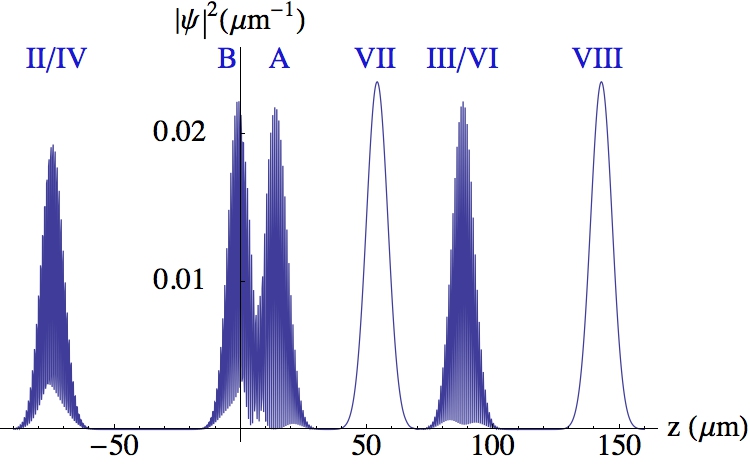}
\caption{\label{fig:pattern}
Numerical simulation of the split electron beam after the last laser
pulse. The offset between beam A and B is a consequence of Lorentz
contraction. See Sec.~\ref{sec:Numeric} for further details.}
\end{center}
\end{figure}
In Fig.~\ref{fig:pattern}  we show a full numerical simulation of the
interference pattern, including the full Kapitza-Dirac effect
as described in  Ref.~\cite{KPM-eInterferometer2013} 
instead of the simplified rule (iv).
We assume fourth-order
harmonic generation ($n=4$), which corresponds to $\beta=3/5$, so
that electrons
need to travel through an electric potential difference
of 127.75~kV (see Fig.~\ref{fig:expSketch}). The flight durations used in the simulation 
are  $T=50$ ns, 
$T'=12.5$ ns, and $\tilde{T}=31$ ns. The distance between 
wavepackets A and B of about 15 $\mu$m corresponds to the spatial
mismatch
induced by Lorentz contraction. Wavepackets with Roman numbers are labeled
in the same way as in Ref.~\cite{KPM-eInterferometer2013} and represent
other partial beams.

The details of the numerical simulation are as follows. We performed
all calculations in momentum space and used a grid of 10000 points
for a Fourier transformation of the final wavefunction to obtain the
spatial density shown in Fig.~\ref{fig:pattern}.
Rules (i) and (ii) can be evaluated exactly in momentum space.
For the acceleration of $1.8\times 10^{16}$~m/s$^2$ that we
considered, rule (iii) provides an excellent approximation and can
also be evaluated directly. To implement an accurate description of
rule (iv) we use the
result that each bi-chromatic pulse couples a wavefunction with
momentum $p$ to wavefunctions with momentum 
$p\pm 2\hbar k_L$ \cite{KPM-eInterferometer2013}. As these
momenta are coupled to other momenta, one obtains a coupling
between an infinite set of momenta separated by multiples
of $2\hbar k_L$. However, for a pulse
duration of $t_L=\pi/(4 g_1 g_2) = 1.56$ ns, where
$g_i$ is defined in Eq.~(7) of Ref.~\cite{KPM-eInterferometer2013}, the laser intensities are so
low that this coupling is very weak and can be neglected except
when resonance condition (\ref{eq:resCond}) 
(or Eq.~(12) of Ref.~\cite{KPM-eInterferometer2013} for
non-relativistic electrons)  
is fulfilled. We therefore only need to
take into account the coupling between two resonant momenta, so that
the unitary time evolution operator during a laser pulse can be found
analytically by diagonalizing a $2\times 2$ matrix. We remark that
this procedure is more accurate than Eq.~(\ref{eq:UtrafoSimpl}), which
neglects the actual time evolution during the laser pulses and only
describes how the electron momentum is changed.

The method presented in the preceding paragraph is non-relativistic
and provides 
an accurate description of
the electron dynamics during the first two laser pulses. To apply this
procedure to the last two laser pulses we performed a Lorentz
transformation of the numerical wavefunction at the end of the
acceleration phase into the (mean) rest
frame of the electrons in beam B, taking into account the changes to the vector
potential discussed in Sec.~\ref{Sec:BS}. The evolution during the last two laser
pulses and the free evolution between these pulses is then
evaluated in the rest frame of beam B. Fig.~\ref{fig:pattern} shows the final
wavefunction in this frame.

To check the numerical results we have verified that the wavefunction is normalized and the
mean position of all partial beams agrees with
the position (in the rest frame of beam B) of a classical relativistic point particle that receives
a specific momentum kick at each laser pulse. We remark that the
distance between wavepackets A and B in Fig.~\ref{fig:pattern} is not
exactly equal to the shift $(\gamma -1) \Delta z $ derived in
Sec.~\ref{sec:expSketch}, but is rather given by
\begin{equation} 
  z_A- z_B = (\gamma -1) \Delta z 
     + \gamma t_L \frac{ 2 \hbar k_L}{m} 
   - t_L \frac{ 1 \hbar \tilde{k}_L}{m} .
\label{eq:dzCorrect}\end{equation} 
The reason is that Eq.~(\ref{eq:dzCorrect}) takes into account the
finite duration $t_L$ of the laser pulses, while the discussion in
Secs.~\ref{sec:expSketch} and \ref{sec:theory} assumes that the momentum of the electrons
changes instantaneously.

\section{Bell's spaceship paradox}\label{sec:Bell}
Bell's spaceship paradox, which was popularized by Bell
\cite{BellSpeakable} but
originally suggested by Dewan and Beran \cite{dewan:517},
is one of the thought experiments 
illustrating the subtleties of Special Relativity. Two
spaceships are initially at rest and connected by a taut thread.
They undergo the same acceleration until they
reach relativistic speed. An observer in the lab frame would
conclude that the thread will not break because the distance between
the spaceships would never change. However, in the reference frame of
the ships Lorentz contraction would imply that the thread
should break.

\begin{figure}
\begin{center}
\includegraphics[height=3.8cm]{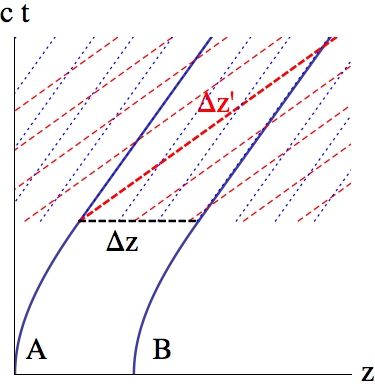}
\hspace{3mm}
\caption{\label{fig:stDiagram}
Space-time diagram of Bell's spaceship paradox. Solid blue lines
correspond to the world line of the two ships. Dashed red lines
(dotted blue lines) 
are the spatial (temporal) coordinate lines, respectively, in the
reference frame 
of the spaceships after they were accelerated.}
\end{center}
\end{figure}

The paradox can be explained using the
space-time diagram shown in Fig.~\ref{fig:stDiagram}.
Blue solid lines describe the trajectories of 
two spaceships (A and B), which are initially
separated by a distance $\Delta z$. They are
accelerated until they reach relativistic speed.
After the acceleration an observer in the lab frame would
measure the distance between the ships simultaneously in her
frame, along the horizontal black dashed line in
Fig.~\ref{fig:stDiagram} a). She would conclude that the distance
is still given by $\Delta z$, which is the
proper distance between the end points of the black dashed line.
An observer on a ship
would measure the distance simultaneously in his frame, 
along the bold dashed red line in the figure. 
The reason for the change in the distance is relativity of
simultaneity. In the reference frame of ship A,
ship B stopped to accelerate
earlier and thus had more
time to travel to the right.

To analyze the paradox, some authors use space-time diagrams only
\cite{Evett1972,0031-9120-40-6-F03,0143-0807-29-3-N02,Matsuda:AAPPS2004}.
Other authors address the question whether Lorentz contraction will 
cause stress forces in the string to occur, which may trigger the
string to break and thus provide a physical signal that resolves the paradox.
The answer to this question is much more
involved due to the subtleties of relativistic rigid-body dynamics
and has been addressed with different methods and results
\cite{dewan:517,BellSpeakable,Nawrocki:AJP1962,Dewan:1962AJP,0295-5075-71-5-699,0143-0807-31-2-006,0143-0807-24-2-361,styer:805}. Most
authors came to the conclusion that the string would break.

Comparing Figs.~\ref{fig:expSketch}b) and \ref{fig:stDiagram} one can
see that Bell's paradox and the proposed interference experiment
are closely related. Except for the parts in which electron beam A and B
are split and recombined, the two space-time diagrams coincide.
In both cases, it is Lorentz contraction of the final separation
$\Delta z'$ that is responsible for a physical effect: the mismatch
between the final positions of beam A and B in the electron
interferometer, and the breaking of the string in Bell's paradox.
One may say that the mismatch in the interference experiment 
replaces the breaking of the string as a physical signature for
the resolution of the paradox.

\section{Discussion}
The theoretical analysis that we have presented above is based
on several simplifying assumptions, including a homogeneous
electric field, laser pulses that are switched on and off at specific
times, and a one-dimensional analysis that ignores forces
in the $x$- and $y$-direction.
In this section we estimate how
deviations from these assumptions may affect the
proposed experiment. In doing so it is important to keep the following points in mind:

(i) Strictly speaking, the proposed experiment is not an interference
experiment. The measured quantity is a displacement between to partial
beams, which would also be produced for incoherent electron beams.
However, if the two partial beams are partially overlapping, the
fringe visibility can provide information about the displacement.
Coherence is therefore helpful but not essential.

(ii) The measured observable is a relative displacement of two partial
beams along the direction of the laser
pulses. Any effects that affect the motion in other directions, such
as forces in the $x$- or $y$-direction, do therefore not affect the
result. Similarly, forces in the $z$-direction that affect both
partial beams in the same way will not affect the displacement.

(iii) The beam splitting process is velocity-selective, i.e., the
electron beam is only split or recombined for electrons within a specific
velocity range.

With these remarks we can address a number of experimentally relevant
questions.

{\em Pulse timing}. In our theoretical analysis the
counter-propagating light pulses are
switched on and off simultaneously everywhere in space. In reality,
the pulses are propagating in opposite directions and will hit the
fast moving partial electron beams at different times. In the proposed
setup this is not a problem because only partial beam A actually
interacts with the pulses, while partial beam B does not obtain a
momentum transfer. In an experiment, pulse timing should therefore
be designed in such a way that partial beam A interacts with the
pulses at the correct time.

{\em Pulse shape}. The spatial shape of a light pulse also affects
the force light exerts on charged particles. Because of point (ii)
above, the transverse pulse profile will not have a significant 
influence on the displacement between beam A and B. As a rule of
thumb, the magnitude
of the momentum change due to the envelope ${\cal E}$ of a pulse is much
smaller than the momentum transfer in a resonant absorption or
emission process as long as the envelope changes slowly over the range
of one wavelength $\lambda$, so that $\lambda |\nabla {\cal E}| \ll |{\cal E}|$
\cite{Wallis1995203}. The transverse profile would therefore only
generate a displacement in the $x$-$y$-plane that is much smaller
than the displacement of partial beam VIII in
Fig.~(\ref{fig:pattern}). Similar remarks apply to the longitudinal
pulse profile. However, the longitudinal profile must be shaped in
such a way that the electron-pulse interaction is 
not switched on adiabatically \cite{Fedorov1974205}.

{\em Photon emission}. It is well known that accelerated charges emit
radiation. In matter-wave interferometry, even the emission of a
single photon may lead to a loss of coherence \cite{PRL73:1223}.
While point (i) implies that this loss of coherence is not a fundamental
problem, the associated change in the electron momentum may
nevertheless affect the displacement between partial beams A and B.
It is therefore worthwhile to estimate the probability of photon
emission during the acceleration.

The photon emission probability for an accelerated electron per unit time and
transverse momentum $p_x,p_y$ is given by \cite{PhysRevD.46.3450}
\begin{align} 
  P(p_x, p_y) &= \frac{  q^2c^2}{\pi^2 a\varepsilon_0 \hbar^3} 
  \left | K_1\left ( \frac{c^2}{\hbar a} \sqrt{p_x^2+p_y^2}\right )\right |^2 ,
\end{align} 
where $K_1$ denotes the modified Bessel function of the second kind.
The total probability to emit a photon can then be estimated by
\begin{align} 
  P_\text{em} &= T' \int dp_x \, dp_y P(p_x, p_y)
\\ &= 2\pi  T' \int_0^\infty  dk_\perp \,
     \frac{  q^2c^2}{\pi^2 a\varepsilon_0 \hbar} 
     \left | K_1\left ( \frac{c^2}{ a} k_\perp\right )\right |^2 ,
\end{align} 
with $k_\perp = \sqrt{p_x^2+p_y^2}/\hbar$ the transverse wavenumber of
the electron. This integral is logarithmically divergent for large
wavelengths, $k_\perp \rightarrow 0$. We regularize it by replacing
the lower boundary by $u\, a/c^2$, where $u>0$ parametrizes the
value of the cutoff
and $c^2/a$ corresponds to the
maximal distance (the largest wavelength) that fits into a Rindler
wedge \cite{BirrellDavies} in the reference frame of an
accelerated observer. The integral can then be performed numerically.
For the parameters used in Sec.~\ref{sec:Numeric},
$P_\text{em}(u)$ varies very slowly with $u$ and is less than 5\%
for $u>10^{-4}$. We therefore expect that photon emission will not
pose a problem for the proposed experiment.

{\em Spatial field fluctuations}. In the previous sections we have assumed
that the electric field is homogeneous and time independent. Spatial
homogeneity is not a critical assumption because it only matters that
both partial beams achieve the same Lorentz factor $\gamma$ at the end
of the acceleration phase. Because both partial beams
essentially follow the same path and are only separated along the
$z$-axis, both beams would undergo the same (non-constant)
acceleration before they are recombined. Hence, field fluctuations
along the $z$-axis would not affect the experimental outcome. Transverse field
fluctuations in the $x$- or $y$-direction could result in different
acceleration for both beams, but this would be accompanied by
a displacement of the beams in the $x$-$y$ plane. For a given point
 in the $x$-$y$ plane, the $z$-displacement should still be the same.
The only spatial field fluctuations that would be of concern are those
which couple transverse and longitudinal motion of the electrons.
They can be dealt with in a similar way as temporal fluctuations (see below).

{\em Temporal field fluctuations}. 
To avoid electric forces between
two partial beams, the experiment should be performed in such a way 
that only one electron passes through the interferometer in each run.
Temporal fluctuations in the electric field could significantly change the
dynamics of the electrons between different runs and thus make it
impossible to measure the beam displacement.
Fortunately,
point (iii) provides a way to overcome this problem: only electrons
that are at the right time, and with the correct velocity,
at the location of the laser pulse will interact with
it. Thus, the resonant interaction with laser pulses selects those
electrons which have obtained the proper velocity and position
to contribute to the measured observable.

In the setup shown in Fig.~\ref{fig:expSketch}, velocity-selection
would only apply to partial beam A, because beam B does not interact
with the laser pulses. The setup could be modified in such a way
that both beams A and B would receive a momentum transfer from
(possibly different) laser pulses after the acceleration. In this
way, both beams would be subject to velocity-selection. Furthermore,
such a modification could be used to move beam A, B away from 
background electrons that do not interact with the laser pulses,
similarly to beam VIII in Fig.~\ref{fig:pattern}.
A disadvantage of velocity-selection is that runs in which an
electron has the wrong velocity
will not contribute to the measurement. The total number of 
experimental runs needed will therefore be increased. 

{\em Detection}. 
To detect the interference pattern of electrons moving at
relativistic speed, a time-of-flight measurement may be
needed. Alternatively, it would be possible to decelerate the
electrons after the last laser pulse and detect the electrons
when they obtained non-relativistic speed. Such a deceleration
phase would lead to a Lorentz contraction of the distance between the
two electron beams, but it would not undo the separation.

\section{Conclusion}
 We have proposed an experiment in which 
Lorentz contraction changes the interference
pattern in an electron interferometer. Two partial beams,
which would be perfectly overlapping for non-relativistic
electrons, will miss each other by an amount
$\Delta z (\gamma -1)$, with $\Delta z$ the beam
separation in the interferometer. The experiment is closely
related to Bell's spaceship paradox, with the mismatch of the
beams replacing the breaking of a string as physical evidence
for Lorentz contraction.

The key element of our proposal is the use of laser pulses
to modify the electron motion via the Kapitza-Dirac effect.
Using fields instead of gratings as beam splitters
enables us to fix time and location of the
splitting process in the 
rest frame of the electrons, rather than in the lab frame. In combination 
with the large accelerations that can be achieved for
charged elementary particles in general, this method
may pave the way 
to further tests of relativity, such as local Lorentz invariance
\cite{0264-9381-11-4-021} or 
extended relativity \cite{1402-4896-82-1-015004}.

\acknowledgments
This project was funded by NSERC, ACEnet and 
UCR of St.~Francis Xavier University. We thank 
J.~Franklin for comments on a previous version of the manuscript.
\\[3mm] 
\hrule
$ $\\[2mm]
\bibliography{/Users/pmarzlin/Documents/literatur/kpmJabRef.bib}
\end{document}